\documentclass[aps,prb,twocolumn,showpacs,showkeys,superscriptaddress]{revtex4-1}

\usepackage{graphicx}% Include figure files
\usepackage{dcolumn}% Align table columns on decimal point
\usepackage{bm}% bold math
\usepackage{multirow}
\usepackage{amsmath,amssymb}
\usepackage{graphicx}
\usepackage{ulem}
\usepackage{color}
\usepackage{lineno}

\definecolor{red}{rgb}{1,0,0}
\definecolor{blue}{rgb}{0,0,1}
\definecolor{black}{rgb}{0,0,0}

\begin{document}

%\preprint{}

%\title{Superconductivity in the topological superconductor candidate La$_{3}$Pt$_{3}$Bi$_{4}$}
\title{Superconductivity in the nodal-line compound La$_3$Pt$_3$Bi$_4$}



\author{Liang Li}
\thanks{These authors contribute equally.}
\affiliation {Hangzhou Key Laboratory of Quantum Matter, School of Physics, Hangzhou Normal University, Hangzhou 311121, China}

\author{Guo-Xiang Zhi}
\thanks{These authors contribute equally.}
\affiliation {Department of Physics, Zhejiang University, Hangzhou 310027, China}

\author{Qinqing Zhu}
\affiliation {Hangzhou Key Laboratory of Quantum Matter, School of Physics, Hangzhou Normal University, Hangzhou 311121, China}

\author{Chunxiang Wu}
\affiliation {Department of Physics, Zhejiang University, Hangzhou 310027, China}

\author{Zhihua Yang}
\affiliation {Hangzhou Key Laboratory of Quantum Matter, School of Physics, Hangzhou Normal University, Hangzhou 311121, China}

\author{Jianhua Du}
\affiliation {Department of Physics, China Jiliang University, Hangzhou 310018, China}

\author{Jinhu Yang}
\affiliation {Hangzhou Key Laboratory of Quantum Matter, School of Physics, Hangzhou Normal University, Hangzhou 311121, China}

\author{Bin Chen}
\affiliation {Hangzhou Key Laboratory of Quantum Matter, School of Physics, Hangzhou Normal University, Hangzhou 311121, China}

\author{Hangdong Wang}
\email{hdwang@hznu.edu.cn}
\affiliation {Hangzhou Key Laboratory of Quantum Matter, School of Physics, Hangzhou Normal University, Hangzhou 311121, China}

\author{Chao Cao}
\email{ccao@hznu.edu.cn}
\affiliation {Department of Physics, Zhejiang University, Hangzhou 310027, China}
\affiliation {Center for Correlated Matter, Zhejiang University, Hangzhou 310058, China}

\author{Minghu Fang}
\email{mhfang@zju.edu.cn}
\affiliation {Department of Physics, Zhejiang University, Hangzhou 310027, China}
\affiliation {Collaborative Innovation Center of Advanced Microstructures,  Nanjing University, Nanjing 210093, China}

%\thanks{}


\date{\today}
%\linenumbers
\begin{abstract}

Owing to the specific topological states in nodal-line semimetals, novel topological superconductivity is expected to emerge in these systems. In this letter, by combination of the first-principles calculations and resistivity, susceptibility and specific heat measurements, we demonstrate that La$_3$Pt$_3$Bi$_4$ is a topologically nontrivial nodal-ring semimetal protected by the gliding-mirror symmetry even in the presence of spin-orbit coupling. Meanwhile, we discover bulk superconductivity with a transition temperature of $\sim$1.1 K, and an upper critical field of $\sim$0.41 T. These findings demonstrate that La$_3$Pt$_3$Bi$_4$ provides a material platform for studying novel superconductivity in the nodal-ring system.

\end{abstract}


\maketitle
%\linenumbers
\section{INTRODUCTION}
Searching for topological superconductors (TSCs) with Majorana fermions has been one of the hottest topics in contemporary condensed matter physics. The discovery of unconventional superconductivity \cite{Y.Hor,T.Kawai,S.Yonezawa,S.Sasaki,R.Tao} in the Cu-intercalated Bi$_{2}$Se$_{3}$, a topological insulator, has initiated intense interest for TSCs. Although the $p$-wave superconductivity is considered to be an intrinsic TSC, in which the core of the vortex contains a localized quasiparticle with exactly zero energy \cite{X.Qi}, the properties of several experimental candidates remain debatable \cite{A.Mackenzie,A.Pustogow}. TSCs can also be realized at the interface of a heterostructure between a strong topological insulator (TI) and an $s$-wave superconductor due to the proximity effect \cite{L.Fu,J.Williams,M.Wang,F.Yang}, where the control of the chemical reaction and lattice mismatch at the interface remains challenging. The most promising candidates for TSCs are those in which fully-gapped bulk superconductivity coexists with topologically protected gapless surface/edge states, such as FeTe$_{1-x}$Se$_{x}$ \cite{M.Fang,Z.Wang}, $\beta$-PdBi$_{2}$ \cite{M.Sakano} and 2M-WS$_{2}$ \cite{Y.Fang}. Recently, nodal-line materials have received significant attention \cite{A.Burkov,H.Weng,T.Bzduvsek}, because they are predicted to exhibit several notable phenomena, including long-ranged Coulomb interaction, large surface polarization charge, $etc$. \cite{Y.Kim,Y.Huh,M.Hirayama,S.Ramamurthy} In particular, if superconductivity is induced in nodal-line materials \cite{G.Bian} with torus-shaped Fermi surfaces (FSs) derived from nodal loops, then topological crystalline and second-order topological superconductivity can be realized \cite{H.Shapourian}. Unfortunately, the nodal-line structure is difficult to realize in materials because it generally become gapped if the spin-orbit coupling (SOC) is considered \cite{C.Fang,M.Ali,A.Yamakage}. So far, nodal-line materials are rarely reported in the presence of SOC \cite{J.Carter,Y.Chen,Q.Liang,Y.Sun}, let alone the superconductors with such nodal lines \cite{G.Bian,A.Ikeda}.

Recently, a possible nontrivial band topology has been reported and enthusiastically discussed based on the strongly correlated Ce$_{3}$(Pt$_{1-x}$Pd$_{x}$)$_{3}$Bi$_{4}$ system \cite{S.Dzsaber,C.Cao}, thereby resulting in renewed focus on the Ln$_3$T$_3$X$_4$ compounds (Ln = lanthanoid element, T = Cu, Au, Rh, Pd, Pt and X = Sb, Bi) \cite{H.Wijn}. As a member of Ln$_3$T$_3$X$_4$ family, La$_3$Pt$_3$Bi$_4$ crystallizes in a cubic Y$_{3}$Au$_{3}$Sb$_{4}$-type structure with space group $I\bar{4}3d$ (No. 220), which contains six gliding-mirror symmetry operations. To perform a comparison on the Kondo insulator Ce$_3$Pt$_3$Bi$_4$, measurements of resistivity, magnetic susceptibility, and specific heat of La$_3$Pt$_3$Bi$_4$ above 2 K were performed, and no superconductivity was reported \cite{G.Kwei,M.Hundley,M.Hundley2,T.Pietrus}. In this letter, we employed the density functional theory (DFT) to investigate the electronic band structure and its band topological properties of La$_3$Pt$_3$Bi$_4$. The results indicate that La$_3$Pt$_3$Bi$_4$ is a topologically nontrivial nodal-ring semimetal. Meanwhile, by performing resistivity, magnetic susceptibility, and specific heat measurements, we discover that La$_3$Pt$_3$Bi$_4$ crystals exhibit bulk superconductivity with a transition temperature $T_{C}$ $\sim$1.1 K and an upper critical field $\mu$$_{0}$$H_{c2}$(0) $\sim$0.41 T. These findings demonstrate that La$_3$Pt$_3$Bi$_4$ provides a material platform for studying novel superconductivity in the nodal-ring system.

\section{EXPERIMENTAL AND COMPUTATIONAL METHODS}

\begin{figure*}
  \includegraphics[width=15cm]{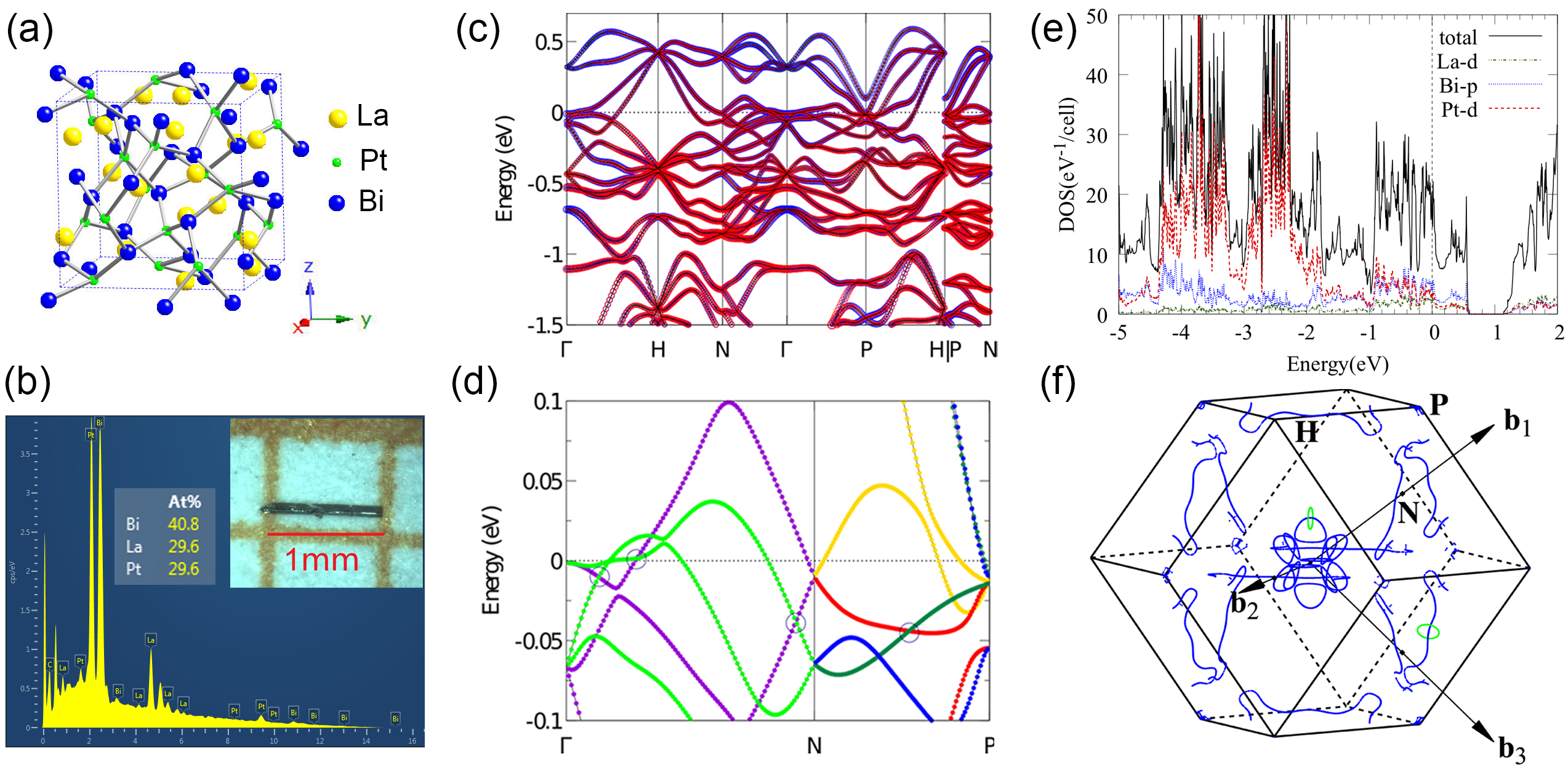}\\
  \caption{(Color online) (a) Crystal structure of La$_{3}$Pt$_{3}$Bi$_{4}$ with a Y$_{3}$Au$_{3}$Sb$_{4}$-type structure (SG: $I\bar{4}$3d); (b) EDX spectrum of a La$_{3}$Pt$_{3}$Bi$_{4}$ single crystal. The inset is a photo of La$_{3}$Pt$_{3}$Bi$_{4}$ crystal; (c) Electronic band structure of La$_3$Pt$_3$Bi$_4$ plotted with orbital characters along high symmetry lines. Red/blue denotes Pt-5$d$/Bi-6$p$ orbitals, and the width of the lines are proportional to the weight of the orbital; (d) Electronic band structure plotted with irreducible representations of the little group along $\Gamma$-$N$ and $N$-$P$. Along $\Gamma$-$N$, the dark violet(green) lines has eigenvalue of -$i$(+$i$) for the gliding-mirror symmetry; while along $N$-$P$, the red lines have eigenvalues -1 and -$i$, gold lines have 1 and -$i$, dark-green lines have 1 and +$i$, and blue lines have -1 and +$i$ for the screw and gliding-mirror symmetries, respectively; (e) PDOS of La$_{3}$Pt$_{3}$Bi$_{4}$ in the -5 to 2 eV energy range relative to the Fermi energy; (f) The first Brillouin zone showing the high symmetry points and the nodal ring structure (blue lines) of La$_3$Pt$_3$Bi$_4$. The green circles are the $K$-path used to calculate the Berry phase.}
\end{figure*}

La$_{3}$Pt$_{3}$Bi$_{4}$ crystals were grown using the method described in Ref. \cite{S.Dzsaber}, and the crystals with typical dimensions of 0.1 $\times$ 0.1 $\times$ 1 mm$^{3}$ were obtained \cite{T.Pietrus}, as shown in the inset of Fig. 1(b). Based on single crystal $x$-ray diffraction (Rigaku Gemini A Ultra), the crystal structure was confirmed to be a cubic [see Fig. 1(a)], with lattice parameters $a$ = $b$ = $c$ = 10.175(5) \AA, which agrees well with the results reported in Ref. \cite{G.Kwei}. Energy-dispersive $x$-ray spectroscopy measurements were performed using a Zeiss Supra 55 scanning electron microscope to verify the crystal composition of La:Pt:Bi = 3:3:4 [see Fig. 1(b)]. The resistivity, and specific heat were measured using a $Quantum Design$ Physical Properties Measurement System (PPMS-9), with a $^{3}$He refrigerator attachment down to 0.5 K. The dc magnetization of the crushed powders was measured using a commercial SQUID magnetometer ($Quantum Design$ MPMS3).

Electronic structure calculations of La$_3$Pt$_3$Bi$_4$ were carried out using plane-wave basis DFT as implemented in the Vienna Abinit Simulation Package (VASP)\cite{method:vasp,method:pawvasp}. The valence-ion interactions were approximated using the projected augmented wave method\cite{method:paw}, and the exchange-correlation functional was approximated using the Perdew, Burke, and Ernzerhoff flavor of the general gradient approximation \cite{method:pbe}. The SOC was considered throughout the calculation as a second variation to the total energy. The plane-wave basis energy cutoff was chosen to be 370 eV, and a $6 \times 6 \times 6$ $\Gamma$-centered $K$-mesh was used for Brillouine zone integration to ensure the convergence of the total energy to 1 meV/atom. The geometry was fully relaxed to all forces $<$ 1 meV/\AA\ , and internal stress $<$ 0.1 kBar. Finally, the electronic structure obtained from VASP was fitted to a tight-binding Hamiltonian formed by the La-5$d$, Pt-5$d$, and Bi-6$p$ orbitals using the Wannier projection method\cite{method:mlwf,method:wannier90}. The symmetrization of the resulting Hamiltonian and symmetry analysis of the band structure were performed using the WannSymm code\cite{ZHI2022108196}. The symmetrized Hamiltonian was then employed to calculate the nodal structure and relevant topological invariants using the WannierTools code\cite{WU2017}.

\section{RESULTS AND DISCUSSION}

\begin{figure*}
  \includegraphics[width=12cm]{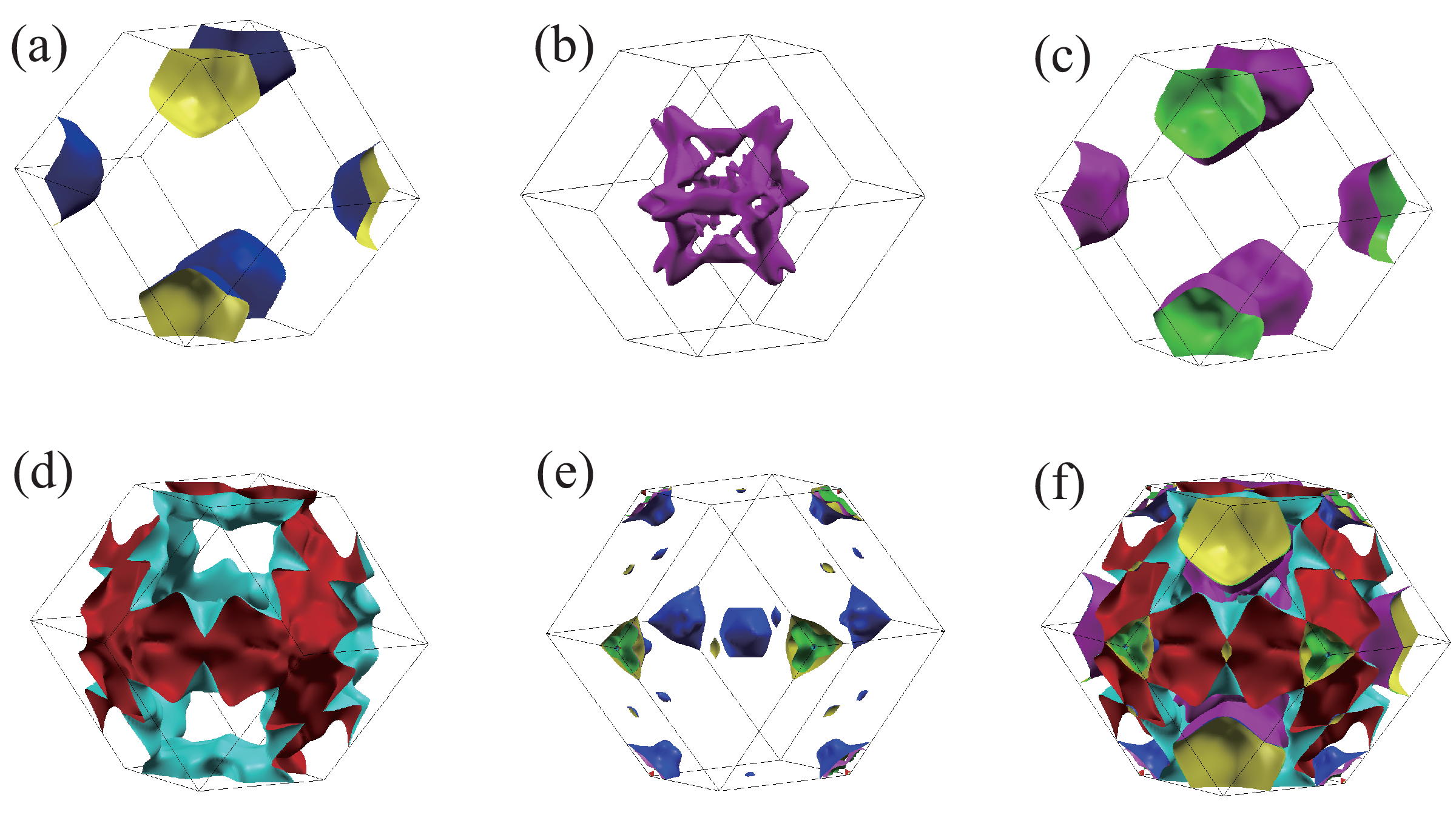}\\
 \caption{(Color online) Fermi surface of La$_3$Pt$_3$Bi$_4$. (a-e): Separated fermi surface sheets. Three of the Fermi surface sheets around P are too small, so we plot them together in panel (e). (f): Full Fermi surface plot. }
  \label{fig:fs}
\end{figure*}

First, we discuss the electronic band structure and its topological nature based on the DFT calculations. Figures 1(c) and 1(d) show the electronic band structure of La$_3$Pt$_3$Bi$_4$ with SOC. The electronic states near the Fermi level are dominated by the Pt-5$d$ and Bi-6$p$ orbitals, as shown by the projected density of states (PDOS) in Figure 1(e). The total density of states (DOS) is $n(E_F) = 6.55$ states/(eV$\cdot$f.u.), or equivalently $\gamma_{b}$ = 15.45 mJ/(mol$\cdot$K$^2$). By comparing to the experimentally observed value (please refer to the experimental results), we conclude that the electronic correlation is well described at the DFT level. Large SOC splitting is expected and observed in the resulting band structure. Assuming local Ce-4$f$ states, the band structure is similar to that of Ce$_3$Pt$_3$Bi$_4$ at higher temperatures, as expected \cite{C.Cao}. At $\Gamma$, the highest occupied states are doubly degenerate $\Gamma_7$, which is extremely close to the Fermi level, and the next highest occupied states are quarterly degenerate $\Gamma_8$ around $E_F-70$ meV. All the states in the $\Gamma$-$N$-$P$ plane can be classified using the eigenstates of the gliding-mirror symmetry. The $\Gamma_7$ state consists of a pair of states with opposite eigenvalues ($\pm i$) of the gliding-mirror symmetry, whereas the $\Gamma_8$ state consists of two pairs. Meanwhile, at point $N$, all the states are doubly degenerate because the system preserves the time-reversal symmetry. These two states bear equal eigenvalues under the gliding-mirror symmetry. Therefore, between $\Gamma$ and $N$, an odd number of band crossings must be formed by states with opposite eigenvalues of the gliding-mirror symmetry. A similar argument can also be applied to $N$-$P$. Furthermore, since these states can be classified using the gliding-mirror symmetry, they must form a loop in the $\Gamma$-$N$-$P$ gliding-mirror plane. Hence, the DFT band structure calculations show that La$_3$Pt$_3$Bi$_4$ is a nodal-ring semimetal protected by the gliding-mirror symmetry, as discussed for the isostructure Ce$_3$Pd$_3$Bi$_4$ in Ref. \cite{C.Cao}. However, we must point out that the nodal rings in La$_3$Pt$_3$Bi$_4$ are different from those in Ce$_3$Pd$_3$Bi$_4$ at low temperatures\cite{C.Cao}. The symmetry protected nodal rings in Ce$_3$Pt$_3$Bi$_4$ are located near $P$ point; whereas they are located around $N$ point and $\Gamma$ point in La$_3$Pt$_3$Bi$_4$. Such difference is related with partially localized nature of Ce-4$f$ states in Ce$_3$Pd$_3$Bi$_4$, effectively corresponding to a different electron fillings.

Using the symmetrized tight-binding Hamiltonian obtained with Wannier functions, we identified the nodal ring structure of La$_3$Pt$_3$Bi$_4$ and calculated the Berry phase around the nodal rings [see Fig. 1(f)]. The nodes of La$_3$Pt$_3$Bi$_4$ can be generally classified into four sets: 12 symmetrically equivalent small nodal rings around $\Gamma$ (type-A), 12 symmetrically equivalent large nodal rings across the border of Brillouin zones around $N$ points (type-B), 12 symmetrically equivalent small nodal rings around $P$ points, and individual high symmetry nodal points. Among them, only type-B is protected by the crystal symmetries, which are less than 50 meV below the Fermi level. Type-A nodal rings are not protected by the crystal symmetries, but are even closer (within 10 meV) to $E_F$ and large in the $K$-space. Therefore, we primarily discuss these two types of nodal rings herein. Using the Wilson loop method, we calculated the Berry phase around both the type-A and type-B nodal rings using the $K$-path, as shown by the green circles in Fig. 1(f). Both nodal rings yield Berry phase of $\pi$, suggesting that both are topologically nontrivial. We note that the bands crossing $E_F$ lead to 8 Fermi surface sheets (Fig. 2), among which the 4 pockets around P (panel e of Fig. 2) and 2 pockets around $H$ (panel a and c of Fig. 2) are not related with the nodal rings. The rest two pockets (panel b and d of Fig. 2) are associated with nodal rings around $\Gamma$ and $N$, respectively.

\begin{figure}
  \includegraphics[width=8.6cm]{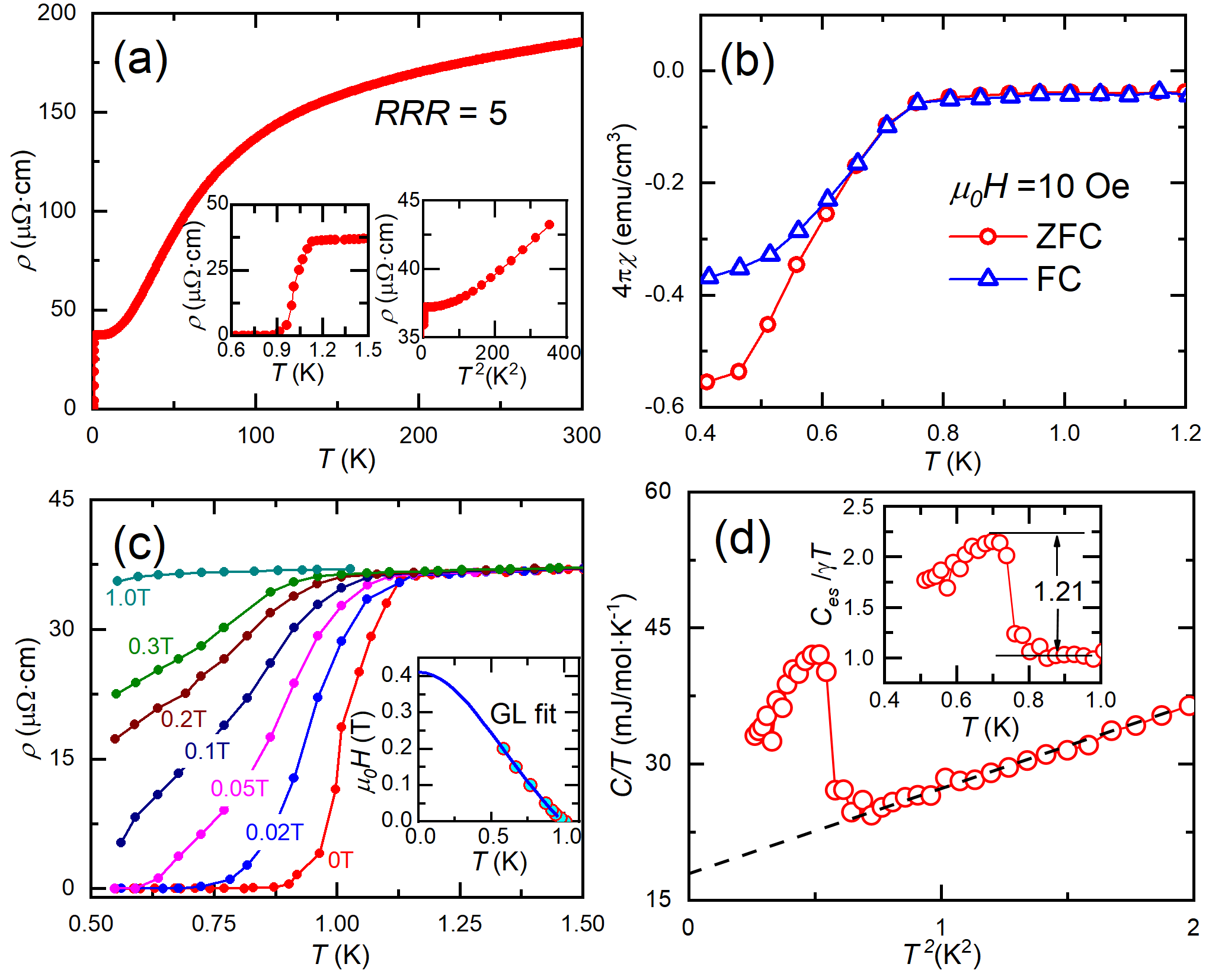}\\
  \caption{(Color online) (a) Temperature dependence of the resistivity of La$_{3}$Pt$_{3}$Bi$_{4}$; Inset: (left) the detail of the superconducting transition; (right) The temperature square dependence of the resistivity below 20 K; (b) $\chi$($T$) near the superconducting transition, measured at 10 Oe field with both ZFC and FC processes; (c) Temperature dependence of the resistivity of La$_{3}$Pt$_{3}$Bi$_{4}$ single crystal measured under various magnetic fields up to 1.0 T; Inset: Temperature dependence of the upper critical field $\mu_{0}H_{c2}$; (d) Specific heat divided by temperature, $C/T$, as a function of $T^{2}$, measured at zero field. The dashed straight line is a guide to the eyes. Inset: Temperature dependence of $C_{es}/\gamma T$ in the superconducting state at zero field, where $C_{es} = C - \beta T^{3}$.}
\end{figure}

Next, we focus on the discovery of superconductivity in La$_{3}$Pt$_{3}$Bi$_{4}$. Figure 3(a) shows the temperature dependence of the resistivity between 0.5 and 300 K for a La$_{3}$Pt$_{3}$Bi$_{4}$ crystal. At $T$ = 300 K, the resistivity is about 185 $\mu\Omega\cdot$cm. Upon cooling down from 300 K, the resistivity exhibits metallic characteristics with a continuous change in the slope. The residual resistivity ratio [$RRR$ = $\rho$(300 K)/$\rho$(2 K)] is about 5, larger than that reported previously \cite{M.Hundley}, indicating the high quality of our crystal. At $T_{C}^{\mathrm{onset}}\sim$1.1 K, the resistivity starts to drop abruptly, then at $T_{C}^{\mathrm{onset}}\sim$ 0.86 K, reaches zero, which is consistent with the onset of a diamagnetic transition [see Fig. 3(b)], indicating the occurrence of a superconducting transition. As shown in the right inset of Fig. 3(a), the temperature dependence of resistivity ($\leq$ 20 K) in the normal state exhibits non-Fermi liquid (NFL) behavior. Since the electronic correlation effect is not so prominent, as evidenced by the comparison between DOS from DFT calculations and $\gamma$, the origin of such NFL behavior is possibly related with the topological nodal ring structure near $E_F$\cite{PhysRevLett.111.206401,PhysRevLett.116.076803}. Figure 3(b) presents the temperature dependence of the susceptibility, $\chi$($T$), below 1.2 K, measured with both zero-field cooling (ZFC) and field cooling (FC) processes. The superconducting volume fraction at 0.4 K is about 37\%, indicating that bulk superconductivity emerges below $T_{C}$ in La$_{3}$Pt$_{3}$Bi$_{4}$. It is worth noting that La$_{3}$Pt$_{3}$Bi$_{4}$ is the first reported superconductor in the Ln$_3$T$_3$X$_4$ (Ln = lanthanoid element, T = Cu, Au, Rh, Pd, Pt and X = Sb, Bi) family, implying that other member may be a superconductor as well; however, this must be confirmed through further investigations. We also note that the half-Heusler LaPtBi compound was found to be a non-centrosymmetric superconductor with a $T_{C}$ of $\sim$0.9 K, $\mu_{0}$$H_{c2}$(0) = 1.5 T, and another candidate for TSC \cite{G.Goll}; however, the superconducting properties, such as $\mu_{0}$$H_{c2}$(0), specific heat jump at $T_{C}$, as well as the properties in the normal state, such as the $\rho$($T$) behavior, differ from those exhibited by La$_{3}$Pt$_{3}$Bi$_{4}$ reported herein.

Figure 3(c) shows the temperature dependence of the resistivity, $\rho$($T,H$), around the superconducting transition measured at various magnetic fields applied perpendicular to the current. With increasing magnetic field, the superconducting transition gradually shifts to a lower temperature. We estimated the $H_{c2}$ using the middle temperature of the superconducting transition ($T_{C}$$^{\mathrm{mid}}$) and plotted $H_{c2}$($T$), as shown in the inset of Fig. 3(c). According to the Ginzburg-Landau (GL) theory, $H_{c2}$($T$) can be fitted using the formula $H_{c2}(T) = H_{c2}(0)(1-t^{2})/(1+t^{2})$, to get the zero-temperature upper critical field $\mu_{0}$$H_{c2}$(0) $\simeq$ 0.41 T, where $t$ is the reduced temperature ($t = T/T_{C}$), as shown by the blue line in the inset of Fig. 3(c). The estimated $H_{c2}$(0) is much lower than that of the half-Heusler LaPtBi compound ($\approx$ 1.5 T) \cite{G.Goll}. Furthermore, the superconducting coherence length $\xi_{0}$ for the La$_{3}$Pt$_{3}$Bi$_{4}$ compound was estimated to be $\sim$28.4 nm using the formula $H_{c2}(0) = \Phi_{0}/2\pi\xi_{0}^{2}$, where $\Phi_{0}$ (2.071 $\times$ 10$^{-15}$ Wb) is the fluxoid quantum.

To measure the specific heat, $C$($T$), we arranged several needle-like crystals on the measurement puck. $C(T)/T$ as a function of $T^{2}$, measured at zero magnetic field, is shown in Fig. 3(d). A significant specific heat jump was observed at approximately 0.8 K, thereby confirming again that La$_{3}$Pt$_{3}$Bi$_{4}$ exhibits bulk superconductivity. We used the formula $C/T = \gamma + \beta T^{2}$ to fit the $C(T)$ data (1.1 - 2 K) in the normal state to obtain the Sommerfeld coefficient $\gamma$ = 17.3 mJ/(mol$\cdot$K$^2$) and the Debye constant $\beta$ = 9.9 mJ/(mol$\cdot$K$^{4}$), corresponding to the Debye temperature $\Theta_{D}$ = 125 K. The electronic specific heat $C_{es}(T)$ in the superconducting state was obtained by subtracting the phonon contribution term $\beta T^{3}$ from the total $C(T)$, as shown in the inset of Fig. 3(d). The normalized specific heat jump $\Delta C/(\gamma T_{C})$ at $T_{C}$ was estimated to be 1.21, which is smaller than the well-known BCS theory value ($\sim$1.43).

Finally, we make a simple analysis of the superconductivity in La$_{3}$Pt$_{3}$Bi$_{4}$ based on our experimental and calculation results. The electronic specific heat coefficient $\gamma$ can be related with $\gamma_b$ via $\gamma=(1+\lambda_{ep})\gamma_b$ in weakly correlated materials, where $\lambda_{ep}$ is the electron-phonon coupling strength. This leads to an estimation of $\lambda_{ep}\approx 0.12$ in this material. Such small electron-phonon coupling results in negligible phonon mediated $T_C$ using BCS formula $T_C=\Theta_D\exp[-1/(\lambda_{ep}-\mu^*)]$, where $\Theta_{D}$ is the Debye temperature, and $\mu^{*}$ is the Coulomb repulsion pseudopotential (chosen within the normal range of 0.10 - 0.15).  Therefore, despite of its low superconducting $T_C$, it is highly possible that the pairing mechanism of La$_3$Pt$_3$Bi$_4$ is unconventional.

\section{CONCLUSIONS}
In summary, by combining the first principles calculations and resistivity, susceptibility and specific heat measurements, we discovered bulk superconductivity with $T_{C}^{\mathrm{onset}}$ = 1.1 K, and $\mu_{0}H_{c2}$(0) = 0.41 T in La$_3$Pt$_3$Bi$_4$ compound. It was confirmed that La$_3$Pt$_3$Bi$_4$ is a nodal-ring semimetal protected by the gliding-mirror symmetry. These results indicate that La$_3$Pt$_3$Bi$_4$ provides a material platform for studying novel superconductivity in the nodal-ring system.

\begin{acknowledgments}
This work was supported by the Ministry of Science and Technology of China under Grant No. 2016YFA0300402 and the National Natural Science Foundation of China (NSFC) (Grant Nos. 11974095, 11874137, 12074335, 11874136), and the Fundamental Research Funds for the Central Universities. The calculations were performed on the High Performance Computing Cluster of Center of Correlated Matters at Zhejiang University, and Tianhe-2 Supercomputing Center.
\end{acknowledgments}

\bibliography{La3Pt3Bi4}

%\begin{thebibliography}{unsrt}

%\end{thebibliography}

\end{document}